\documentclass[conference, 10pt]{IEEEtran}
\IEEEoverridecommandlockouts
\usepackage[noadjust]{cite}
\usepackage{amsmath,amssymb,amsfonts}
\usepackage{algorithm}
\usepackage{algorithmic}
\usepackage{graphicx}
\usepackage{textcomp}
\usepackage{xcolor} 
\usepackage{tikz}
\usepackage{subcaption}
 \usepackage{optidef}
 \usepackage{soul}
\graphicspath{{figures/}}

\usetikzlibrary{shapes,arrows,shadows,positioning}
\usetikzlibrary{arrows, arrows.meta, shapes, shapes.multipart, trees, positioning,
	backgrounds, fit, matrix, petri, calc, automata, chains}
\usepackage{ellipsis}
\usetikzlibrary{calc}
\usetikzlibrary{decorations.pathreplacing,decorations.markings,shapes.geometric}
\usetikzlibrary{patterns}
\def\BibTeX{{\rm B\kern-.05em{\sc i\kern-.025em b}\kern-.08em
    T\kern-.1667em\lower.7ex\hbox{E}\kern-.125emX}}

\author{
	\IEEEauthorblockN{Alperen G\"undo\u{g}an\IEEEauthorrefmark{1}, H. Murat G\"ursu\IEEEauthorrefmark{1}, Volker Pauli\IEEEauthorrefmark{2}, Wolfgang Kellerer\IEEEauthorrefmark{1}} \\
	\IEEEauthorblockA{\IEEEauthorrefmark{1}Chair of Communication Networks, Technical University of Munich}\\
	\IEEEauthorblockA{\IEEEauthorrefmark{2}Nomor Research}\\
		\IEEEauthorblockA{\{alperen.guendogan, murat.guersu, wolfgang.kellerer\}@tum.de, pauli@nomor.de}
}

\title{Distributed Resource Allocation with Multi-Agent Deep Reinforcement Learning for 5G-V2V Communication}

\definecolor{myred}{RGB}{220,43,25}
\definecolor{mygreen}{RGB}{0,146,64}
\definecolor{myblue}{RGB}{0,143,224}
\definecolor{mygray}{gray}{0.80}
\definecolor{mylightergray}{gray}{0.87}
\definecolor{mylightestgray}{gray}{0.95}

\newcommand*{\rom}[1]{\expandafter\@slowromancap\romannumeral #1@}

\begin{document}

\maketitle

\begin{abstract}
We consider the distributed resource selection problem in Vehicle-to-vehicle (V2V) communication in the absence of a base station. Each vehicle autonomously selects transmission resources from a pool of shared resources to disseminate Cooperative Awareness Messages (CAMs). This is a consensus problem where each vehicle has to select a unique resource. The problem becomes more challenging when---due to mobility---the number of vehicles in vicinity of each other is changing dynamically. In a congested scenario, allocation of unique resources for each vehicle becomes infeasible and a congested resource allocation strategy has to be developed. The standardized approach in 5G, namely semi-persistent scheduling (SPS) suffers from effects caused by spatial distribution of the vehicles. In our approach, we turn this into an advantage. We propose a novel DIstributed Resource Allocation mechanism using multi-agent reinforcement Learning (DIRAL) which builds on a unique state representation. One challenging issue is to cope with the non-stationarity introduced by concurrently learning agents which causes convergence problems in multi-agent learning systems. We aimed to tackle non-stationarity with unique state representation.
Specifically, we deploy view-based positional distribution as a state representation to tackle non-stationarity and perform complex joint behavior in a distributed fashion.
Our results showed that DIRAL improves PRR by 20\%  compared to SPS in challenging congested scenarios.
\end{abstract}
\section{Introduction}
Vehicle-to-vehicle (V2V) Communication is a vital technology for automotive industry to reduce the accident risk and to provide safer driving experience. Vehicles periodically broadcast Cooperative Awareness Messages (CAMs) which contain the position, velocity, direction of the vehicles along with emergency vehicle and collision risk warnings~\cite{Communications2018}. 
The 3rd Generation Partnership Project (3GPP) has standardized the message exchange among the vehicles (V2V), and among vehicles and pedestrian (V2P), infrastructure (V2I) and network (V2N) in release~14 \cite{3gpp:36.300}, known as Long-Term-Evolution Vehicle-to-Everything (LTE-V2X). Within 3GPP, the evolution of V2X communication is continuing in the scope of New Radio (NR), a technology for the 5th generation of cellular networks. The allocation of V2V resources in cellular, i.e., time and frequency can be either controlled by the cellular network structure or performed autonomously by the individual vehicles. Since the existence of base stations cannot always be guaranteed, distributed resource allocation (DRA) methods are required. 3GPP has standardized a DRA method that relies on UEs to independently perform sensing and resource selection based on the principle of semi-persistent scheduling (SPS). Based on these two principles vehicles in mutual vicinity are likely to choose the same resources and interfere with each other for a number of subsequent transmissions~\cite{nomor_sps}, resulting in repeated undecoded CAMs sent by these vehicles. Especially, for congested scenarios, it becomes more challenging for the SPS algorithm. 

Our goal is to overcome the drawback of the SPS approach by deriving a DRA that builds on principles otherwise used by centralized resource allocation methodology, namely knowledge of the spatial distribution of vehicles. Achieving the globally optimal solution in DRA is a combinatorial optimization problem and mathematically intractable as the network size increases~\cite{4155374}. Therefore, as a strong heuristic, we adopt the multi-agent deep reinforcement learning (MADRL) approach, i.e., each vehicle is a learning unit. 

MADRL has been recently exploited in various modern network problems such as power allocation; IoT, UAV, and V2X for spectrum access, data rate selection, transmit power control, etc., c.f. \cite{Nasir,8807386,DBLP:journals/corr/abs-1905-02910}. In~\cite{DBLP:journals/corr/abs-1905-02910}, the authors propose MADRL to maximize the delivery rate of V2V messages for in-coverage scenario. They train multiple independent deep Q-networks (DQN) for each V2V link.  As the increase in number of vehicles increases the number of V2V links combinatorically, this method does not scale computationally \cite{Gupta}. We group V2V decisions of each vehicle as a single decision for each vehicle overcoming this scaling problem. Furthermore, we train only one model which is shared by the vehicles such that the vehicles can also learn from the experiences of the other vehicles.

\section{Scenario}
\label{sec:network_model}
\subsection{System Model}
We consider a wireless vehicle communication network consisting of a set of vehicles $ \mathcal{N} = \{1, 2, ..., N\}$.

Vehicles are mobile along a road and their distance to each vehicle is changing dynamically. None of the vehicles are connected to a base-station. Each vehicle has a half-duplex radio, thus only when they are not transmitting, they can receive. For transmission, vehicles autonomously select a resource from a set of available resources $ \mathcal{K} = \{1, 2, ..., K\}$.
For simplification of our system model and to avoid in-band emission (IBE) caused by the simultaneous transmission on adjoining frequencies which degrades system performance in V2X \cite{Bazzi2019}, resources are separated only over time into slots. We consider a time-slotted system in which all vehicles are scheduled simultaneously at the time $t$ for the allocation of resources in $\mathcal{K}$. In this work, we strive to maintain high reliability of periodic broadcast V2V messages i.e. CAMs, in other words, each vehicle is trying to maximize the number of neighbors that decodes its packet. Thus, we use a reliability metric, the packet reception ratio (PRR) from 3GPP \cite{3gpp:36.885} as a key performance indicator. PRR is defined as the ratio of the successful receptions among the total number of neighbors $N_t^i$ of the transmitter vehicle $i$ at time $t$. 
 
Whether a vehicle has successfully received a CAM at a resource depends on if the vehicles in vicinity have selected that resource for transmission. Thus, the selection of resources can be used to model PRR. Each vehicle $i \in \mathcal{N}$ selects an action $a_t^i = k$ which indicates selection of the resource $k\in \mathcal{K}$ at time $t$. The actions selected by all the vehicles at time $t$ is $    \mathbf{a_t} = (a_t^1, a_t^2, ..., a_t^N)$.

For the statement of the problem and the design of the reward we assume a simplified channel model that only reflects path loss, but neglects fast and slow fading. Interference modeling is limited to transmissions from other vehicles.

Note that, if more than one vehicle transmit at the same resource, a receiver only selects the one with the highest SINR for decoding.
Based on this, each vehicle $i$ calculates $\text{PRR}_t^i(\mathbf{a_t})$ as the ratio of the neighbors that correctly decoded its packet at time $t$ given the actions of all vehicles $\mathbf{a_t}$, at time $t$: 
\vspace{-0.1cm}
\begin{equation}
\begin{split}
    \text{PRR}_t^i(\mathbf{a_t}) = \frac{1}{N_t^i} \sum_{j=1}^{N_t^i} 1 \{ P_{err}(\gamma_t^{i,j})\leq X \sim U([0,1])\},
    \end{split}
\vspace{-0.7cm}
\end{equation}
\vspace{-0.1cm}
where $N_t^i$ denotes the number of the vehicles neighbors of the vehicle $i$ at the time $t$, and $\gamma_t^{i,j}$ is the signal-to-interference-plus-noise ratio (SINR) at vehicle $j \in \mathcal{N}_t^i=\{1,\cdots,{N}_t^i \}$ of the packet of vehicle $i$,
\vspace{-0.2cm}
\begin{align}
    \gamma_t^{i,j} = \frac{P {\lvert H_t^{i,j}\rvert}^2}{\sigma^2 + \sum_{k \in N_t^c(\mathbf{a_t})\setminus \{ i \}} P {\lvert H_t^{k,j} \lvert}^2}. \nonumber 
\vspace{-0.3cm}
\end{align}
$P_{err}(\gamma_t^{i,j})$ function calculates the block error rate for the given SINR based on the fixed modulation and coding scheme(MCS). If $P_{err}(\gamma_t^{i,j})$ equal or less than a random number between $0$ and $1$, the packet will be decoded successfully
$P$ is the transmit power that is fixed for all vehicles, $\sigma^2$ is the power of additive white Gaussian noise, $\mathcal{N}_t^c(\mathbf{a_t})$ is the set of interfering packets determined by the actions at the time $t$, $H_t^{i,j}$ and $H_t^{k,j}$ are the channel gain between the transmitter $i$ and receiver $j$ and between the transmitter $k$ and receiver $j$ respectively.
\subsection{Problem Definition}
Vehicles aim to maximize the number of neighbours that receive their CAM messages measured by $\text{PRR}$. 
Given the definition of $\text{PRR}$, we formulate an optimization problem to select the policy $\pi$ that sets the actions of the vehicles $\mathbf{a_t^\pi}$ and maximize the $\text{PRR}$,
\begin{maxi}|1| 
{\mathbf{\pi}}{\sum_{i=1}^{N} \text{PRR}_t^i(\mathbf{a_t^\pi})}
{}{}
\addConstraint{a_t^i \in \{1, 2, ..., K\}.}
\label{eq:prr_sum}
\end{maxi}

For the non-congested case, i.e., $N \leq K $, \eqref{eq:prr_sum} can be satisfied easily if all the vehicles choose separate resources, i.e, $a_t^i~\neq~a_t^l \,\,\forall \,\,i,l \,\,\text{with}$ $i\neq l$. However, with the congestion case, i.e., $N > K$, a policy has to be used to dynamically adjust the actions as the vehicle channel gains and the number of neighbors vary due to the mobility of vehicles. The policy that would maximize the average PRR for all vehicles is called the optimal policy $\pi^*$.

Obtaining the optimal policy $\pi^*$ is challenging for distributed scenario as there is no central scheduler that takes into account the channel gains and locations of each vehicle with respect to each other.  
In centralized solution, the base station coordinates transmission of vehicles, adopts spatial reuse of resources to maintain reliable communications, i.e. high PRR. Specifically, base station allocates the same resources to the vehicles if the distance between them is higher than a minimum reuse distance $r_{reuse}$ \cite{8275637}. However, we consider the case where the vehicles are located outside of the coverage of a base station. Thus, each vehicle selects an action based only on local observations autonomously.

Due to the dynamic multi-dimensional nature of the problem, we formulate
it as a multi-agent deep reinforcement learning problem such that far vehicles ($>r_{reuse}$) are motivated to choose the same resources whereas near vehicles use separate resources. Simply, we investigate the concept of centralized scheduling in a distributed fashion to maximize the overall packet reception ratio.
\subsection{Deep Reinforcement Learning Background}
Reinforcement learning is a machine learning technique where an algorithm, considered as an agent, learns based on its interactions with the environment. Let $\mathcal{S}$ and $\mathcal{A}$ be the state and action space respectively. In a single-agent, fully-observable, RL setting \cite{Sutton1998}, an agent observes the current state $s_t$ in state space $\mathcal{S}$ at each discrete time step $t$, and chooses an action $a_t$ in action space $\mathcal{A}$ based on a policy $\pi$. Then, it observes a reward signal $r_t$, and transitions to a new state $s_{t+1}$. The goal of the agent is to maximize accumulated discounted reward  $G_t:= r_t + \gamma r_{t+1} + \gamma^2 r_{t+2} ...  
= r_t + \gamma G_{t+1}$, where $\gamma \in [0,1]$ is a discount factor which determines the influence of future rewards on the optimal decisions. A policy $\pi$ defines the behaviour of an agent and is the probability of an action given a state, i.e., $\pi(a|s) = \mathbb{P}[a_t = a | s_t = s] $ with $\sum_{a \in \mathcal{A}}\pi(a|s) = 1$. Then following policy $\pi$, the expected total reward starting from the state $s$, taking action $a$ can be calculated via the action-value function $Q^{\pi}(s,a) = \mathop{\mathbb{E}_{\pi}}[G_t|s_t = s, a_t = a]$. The optimal action-value function, denoted as $Q^{*}(s,a)$ is the maximum action-value function over all policies, i.e. $Q^{*}(s,a) = \max_{\pi} Q^{\pi}(s,a)$. Once the optimal action-value function is obtained, the optimal policy $\pi^{*}(s,a)$ can be extracted by acting greedy at each state i.e. $ \underset{a \in A}{\arg\max}\,\, Q^{*}(s,a)$.

One of the most famous algorithm to compute the optimal action-value function is Q-learning \cite{q_learning}, which iteratively approximates the Q-function using the Bellman equation,
\vspace{-0.2cm}
\begin{equation} \label{eq:q_learning}
        Q(s,a) = Q(s,a) + \alpha \big[ r + \gamma \max_{a'}Q(s',a') - Q(s,a)\big].
\vspace{-0.2cm}
\end{equation}
When state and action space is large,  deep neural networks (DNN) are used to approximate the Q function. This technique is known as Deep Q Networks (DQN) where the Q-function can be represented as $Q(s,a;\theta)$ where $\theta$ denotes the trainable weights of the network. In order to find $Q(s,a;\theta)$, the least squares loss $L(\theta)$ is defined;
\vspace{-0.1cm}
\begin{equation} \label{eq:loss}
       L(\theta) = \big[(r + \gamma \max_{a'} Q(s',a'; \theta)) - Q(s,a; \theta) \big]^2.
\vspace{-0.1cm}
\end{equation}
The gradient descent is applied with respect to $\theta$ in order to minimize the loss in (\ref{eq:loss}).
\subsection{Multi-agent DRL Formulation}
We consider a multi-agent system where each reinforcement learning agent, vehicle, learn simultaneously which resources to select for its transmission. At each time-step $t$, each agent $i\,\,\in\,\,\mathcal{N}$ observes a state $s_t^i$ locally, selects an action $a_t^i$ from its policy $\pi^i(a_t^i |s_t^i)$ and receives a reward $r_t^i$ from the environment. The sum of discounted rewards for the agent $i$ for episodes of length $H$ is $G_t^i = \sum_{l=0}^{H} \gamma^l r_{t+l}^i$. 
Each vehicle $i$ aims to find a policy $\pi^i$ to maximize its expected accumulated discounted reward. Note that, the reward of a vehicle $i$ depends not only on the policy $\pi^i$ but also on the policy of the other vehicles. 
The set of policy of all agents except the agent $i$ is denoted by  $\pmb{\pi}^{-i}= \{\pi^{j}\}_{j\neq i}$. We use the shorthand notation $\mathbf{-i}= \mathcal{N} \setminus {i} $ for the set of opponents of agent $i$. Then, the objective of each vehicle $i$ is; 
\vspace{-0.22cm}
\begin{equation} \label{eq:ma_reward}
    \max_{\pi^i} \mathbb{E}\big[G_t^i (\pi^i, \pmb{\pi}^{-i})\big],
\vspace{-0.1cm}
\end{equation}
where $\mathbb{E}\big[G_t^i (\pi^i, \pmb{\pi}^{-i})\big]$ indicates the expected accumulated reward when the agent $i$ follows the policy $\pi^i$ and the opponents perform $\pmb{\pi}^{-i}$. In short, the agents need to learn a cooperative  behaviour in a distributed fashion in order to maximize their objective.
Solving \eqref{eq:ma_reward} is challenging since multiple agents learn concurrently makes the environment \textit{non-stationary} from the perspective of each agent and breaks the stationarity assumption which is required for convergence of single-agent DRL algorithms \cite{zhang2019multiagent}.
One way to deal with \textit{non-stationary} behaviour in multi-agent systems is to anticipate the actions of the other agents through recursive reasoning \cite{hern2017survey}.  A particular case where the agents possess knowledge after infinitive reasoning steps is known as \textit{common knowledge} \cite{10.2307/2958591}. An event, say $\mathrm{e}$, is common knowledge "if and only if everyone knows $\mathrm{e}$, and everyone knows that everyone knows $\mathrm{e}$, and everyone knows that everyone knows that everyone knows $\mathrm{e}$, and so on \textit{ad infinitum}" \cite{Gmytrasiewicz}. 
\section{Multi-agent DRL Algorithm}
\label{sec:model}
\begin{figure}
    \centering
    \includegraphics[width=\textwidth/2]{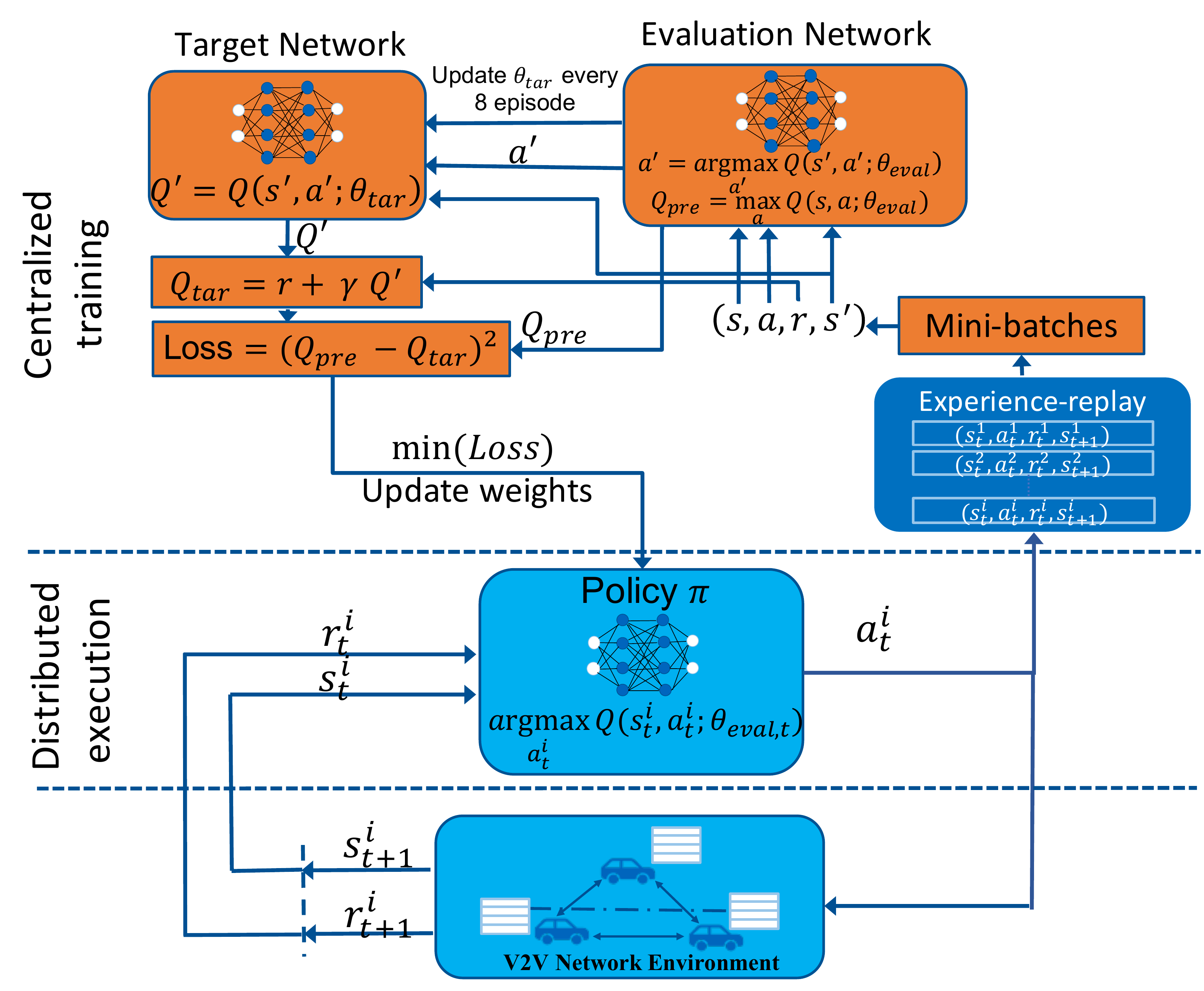}
    \caption{Illustration of DIRAL.}
    \label{fig:diral}
\end{figure}

In this section we describe the solution we propose to the problem~\eqref{eq:ma_reward} that is the DIstributed Resource Allocation with multi-agent deep reinforcement Learning (DIRAL) algorithm.
The novelty of this paper lies in the unique state representation to tackle the \textit{non-stationarity} in multi-agent learning system to perform distributed resource allocation. Each vehicle observes the positions of the other vehicles on the road from its own perspective and creates view-based positional distribution vector as shown in Figure \ref{fig:exp_dist} for vehicle A and C. In this work, \textit{field-of-view common knowledge} \cite{DeWitt2018} arises through view-based positional distribution based observations since vehicles can deduce the observation of other vehicles from their own observation. For example, in Figure \ref{fig:exp_dist}, vehicle A can infer the observation of C from its own observations so that it can allocate a reasonable resource for transmission of CAM by anticipating the resource selection of vehicle C. Cooperative nature of the objective i.e. vehicles need to cooperate for resource allocation to maximize the equation~\eqref{eq:ma_reward}, and centralized training enables vehicles to develop fully decentralized policies under common knowledge \cite{DeWitt2018}.
Figure \ref{fig:diral} depicts the general structure of DIRAL. We focus on centralized training, decentralized execution framework which is commonly used in many MADRL system \cite{Naparstek2017}.
In the centralized training part, we can access the actions of the vehicles to determine the reward for each agent. Once the training part is done, agents exploit the trained policy to take a decision based on only their local observations. Centralized training also facilitates \textit{parameter sharing} approach, such that the parameters of DQN are shared with all the agents. Note that, our current settings allow sharing parameters since agents are homogeneous \cite{Gupta}, i.e. share the same reward utility, state and action space. \textit{Parameter sharing} reduces the number of parameters that must be trained significantly thus training is computationally favourable and scalable. 

We deploy further improvements in order to stabilize the learning and improve the policy. Thus, we adopt double DQN, an enhanced version of DQN with target and evaluation networks to solve the overestimation problem of DQN \cite{DBLP:journals/corr/HasseltGS15}. Evaluation network is used both for action selection and policy evaluation. Target network is exploited to calculate the Q values of the next action i.e. $a\prime$ for computing the loss and is updated with the parameters of evaluation networks periodically. Furthermore, we store the experiences of each agent i.e. $e_t^i = (s_t^i, a_t^i, r_t^i, s_{t+1}^i)$ in the experience replay memory to be exploited for training as depicted in Fig. \ref{fig:diral}. The correlations among experiences can be removed via sampling randomly from the replay memory $(s, a, r, s') \sim U(D)$ during training and the changes in the data distribution smoothing over \cite{mnih2015humanlevel}. We shortened the size of the experience replay and keep a FIFO buffer with a size proportional to the number of agents to avoid non-stationarity introduced in training with large experience replay buffer \cite{leibo2017multiagent}. 

As a part of the deep neural network architecture of DQN, we use  long-term short memory (LSTM) networks as a first layer to predict the mobility pattern of the vehicles based on positional distribution. LSTM layer sustains an internal state and combines the observations over time. We approximate $Q(s_t, a_t, h_{t-1}; \theta)$ with recurrent neural networks where $h_{t-1}$ is the hidden state of the agent at the previous step. The hidden state $h_t = LSTM(s_t, h_{t-1}) = LSTM(s_{t-(L-1)},...,o_t)$ with $L$ as the number of observations. LSTM network is followed by a fully connected feed-forward network layer to compute the values of each action. Note that, although we train only one DQN, the agents still act dissimilar, because each agent evolves its own hidden state due to different observations which enable agents to behave distinctly although they share the same DQN \cite{DBLP:journals/corr/FoersterAFW16a}. 
\subsection{State and Action Space}
The state vector $s_t^i$ at time $t$ for the vehicle $i$ is composed of the previous action that agent $i$ took $a_{t-1}^i$ and the vector represents the positional distribution of other vehicles from the perspective of the vehicle $v_t^i$, 
that is the output of function $f(p_t^i, B, R)$.

\begin{figure}
    \centering
    \includegraphics[width=0.5\textwidth]{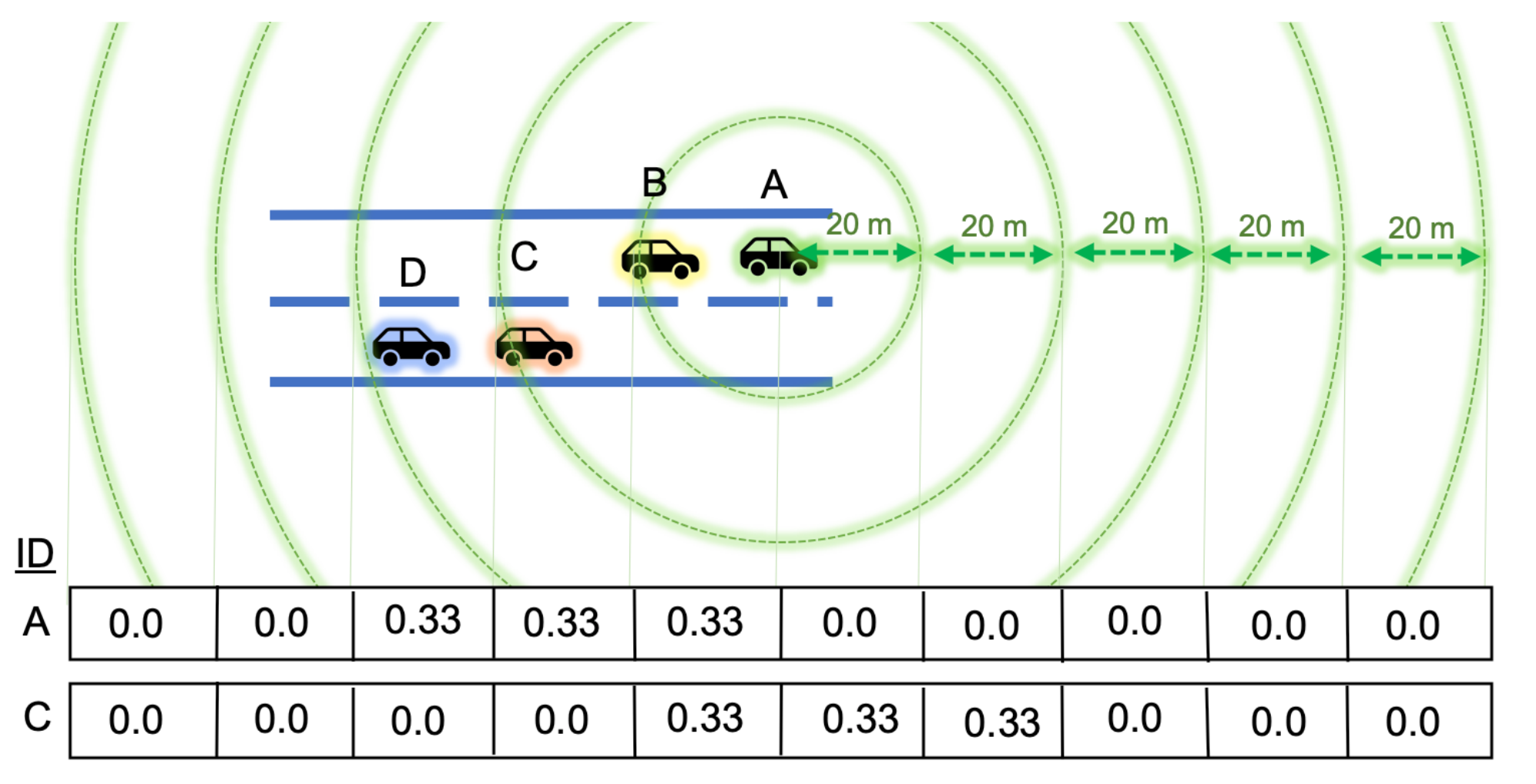}
    \caption{Example of the output of $f(p_t^i, B, R)$ with $B=10$ and $R=100m$}
    \label{fig:exp_dist}
\end{figure}

View-based positional distribution (VPD) function $f(p_t^i, B, R)$  exploits the positions of the other vehicles from the neighbor table $p_t^i$ of the agent $i$ at the time $t$, an integer $B \in \mathbb{Z}^+ $ which determines the granularity of the view-based observation vector and an integer $R \in \mathbb{Z}^+ $ which indicates the observation radius of the agent $i$. The Figure \ref{fig:exp_dist} illustrates the positional distribution of the vehicles $p_t^i$ at the top and the related output of $f(p_t^i, B, R)$ for vehicles A and C at the bottom. The parameters are set as $B=10$ and $R=100m$. The effect of setting these parameters is discussed in Sec.~\eqref{sec:evaluation} with the convergence versus complexity trade-off.  The current neighboring table is piggybacked to the CAM messages as such every vehicle shares their position information. 

In this work, the agent uses all the available frequency chunks with transmissions so that 1 slot and 1 subchannel represent one resource block. 
The action space becomes $\mathcal{A}:= \{a | a = k,\,\, k\in \mathcal{K}\}$.


\subsection{Reward Design}
We consider the performance of the proposed approach both in congestion and non-congestion case. In particular, we encourage each vehicle to select a different resource but for congestion case we motivate far vehicles to use the same resource. The reward of each agent is calculated as follows; 
\begin{equation}\label{eq:reward_design}
    r_t^i(a_t^i|s_t^i) = 
    \begin{cases}
    1, & N_t^c = 1 \\
    \begin{rcases}      
    0 & \text{if $dist(\pmb{c}) > r_{reuse}$}  \\
    -N_t^c & \text{else,} \\
    \end{rcases} &   \text{$N_t^c = 2$} \\
    -N_t^c , & N_t^c > 2 
    \end{cases}
\end{equation}
where the vector $\pmb{c} $ is the interfering agents at the same resource i.e. $\pmb{c} = [i, k,...,l]$ with $\pmb{c} \subset N$ and $N_t^c = |\pmb{c}| $ is the total number of collided vehicles. The average of the sum rewards at the time $t$ is added to the reward of the individual vehicles to intensify cooperative behaviour $    r_t^i(a_t^i|s_t^i) = r_t^i(a_t^i|s_t^i) + \frac{\sum_{j=1}^{N}r_t^j(a_t^j|s_t^j)}{N}$. Note that, we do not need additional feedback channels to inform whether a transmitted packet is successfully decoded or not to compute the reward. We train the model based on only the positional distribution and resource allocation of the vehicles.
\section{Evaluation}
\label{sec:evaluation}
We first evaluate the performance of the proposed approach in a light-weighted test simulator for faster analysis of various experiments. Once the proper architecture is determined in the test simulator, it is deployed and tested in the 5G real-time network simulator (RealNeS)\footnote{http://nomor.de/services/simulation/system-level-simulation/} for evaluation of packet reception ratio. We consider a simple channel model in the test simulator such that if there is more than one vehicle that exploits the same resource block within the range of receivers, receivers decode the packets of the closer vehicle. However, the simulation granularity in RealNeS allows us to compute effective SINRs in the equivalent complex baseband (ECB) taking into account, fast and slow fading, precoding and receive filtering techniques. Table~\ref{tab:parameters} summarizes training and network settings. We evaluated the performance of the DIRAL in the scenarios as shown in Table~\ref{tab:scenarios} for different configurations in terms of number of vehicles and available resources.

\begin{table}[htbp]
\caption{Training and network parameter settings}
\begin{center}
\resizebox{0.5\textwidth}{!}{
\begin{tabular}{|c|c|c|c|}
\hline
\multicolumn{2}{|c|}{\textbf{Training}}&\multicolumn{2}{|c|}{\textbf{Network}} \\
\cline{1-4} 
\textbf{Name} & \textbf{Value}& \textbf{Name}& \textbf{Value} \\
\hline
Time-steps  & 250000  & Access scheme & OFDMA \\
\hline
Experience-replay & 1024 & Carrier frequency & 5.9GHz \\
\hline
Step size(for LSTM)      & 6 & System bandwidth & 10MHz \\
\hline
Batch size(Training)     & 512 & Subcarrier spacing & 30kHz  \\ 
\hline
Learning rate            & 0.0001 &  PRBs & 24 \\ 
\hline
Discount factor $\gamma$ & 0.7 & Number of antennas & 4  \\ 
\hline
Hidden layers            & 256 neurons & Transmit power & 23.0dBm\\
\hline
Optimizer & ADAM & TFC index & 4(QPSK) \\ 
\hline
Activation function & ReLU & CAM message size & 300bytes \\
\hline
$\epsilon$-decay & 0.999 &  Periodicity & 100ms \\
\hline
\end{tabular}}
\label{tab:parameters}
\end{center}
\end{table}
\vspace{-0.65cm}
\begin{table}[htbp]
\caption{Evaluation scenarios}
\begin{center}
\resizebox{0.5\textwidth}{!}{
\begin{tabular}{|c|c|c|c|c|c|}
\hline
\textbf{No} & \textbf{Vehicles} & \textbf{Resources} & \textbf{Highway} & \textbf{Velocity(kmph)}   & \textbf{Mobility} \\
\hline
1           & 4                 & 3                  & 100m                    & \{18,36,45,54\} & Wrap-around       \\ \hline
2           & 6                 & 5                 & 250m                    & $\sim$35       & SUMO$^{\mathrm{1}}$             \\ \hline
3           & 8                 & 10                 & 500m                    & $\sim$35       & SUMO            \\ \hline
4           & 10                & 10                 & 500m                    & $\sim$35        & SUMO              \\ \hline
5           & 12                & 10                 & 500m                    & $\sim$35        & SUMO              \\ \hline
\multicolumn{4}{l}{$^{\mathrm{1}}$https://sumo.dlr.de/docs/.}
\vspace{-0.5cm}
\end{tabular}}
\label{tab:scenarios}
\end{center}
\end{table}
\vspace{-0.2cm}
\subsection{Training Performance}
\begin{figure}
 \centering
\begin{subfigure}{.23\textwidth}
  \centering
  \includegraphics[width=\linewidth]{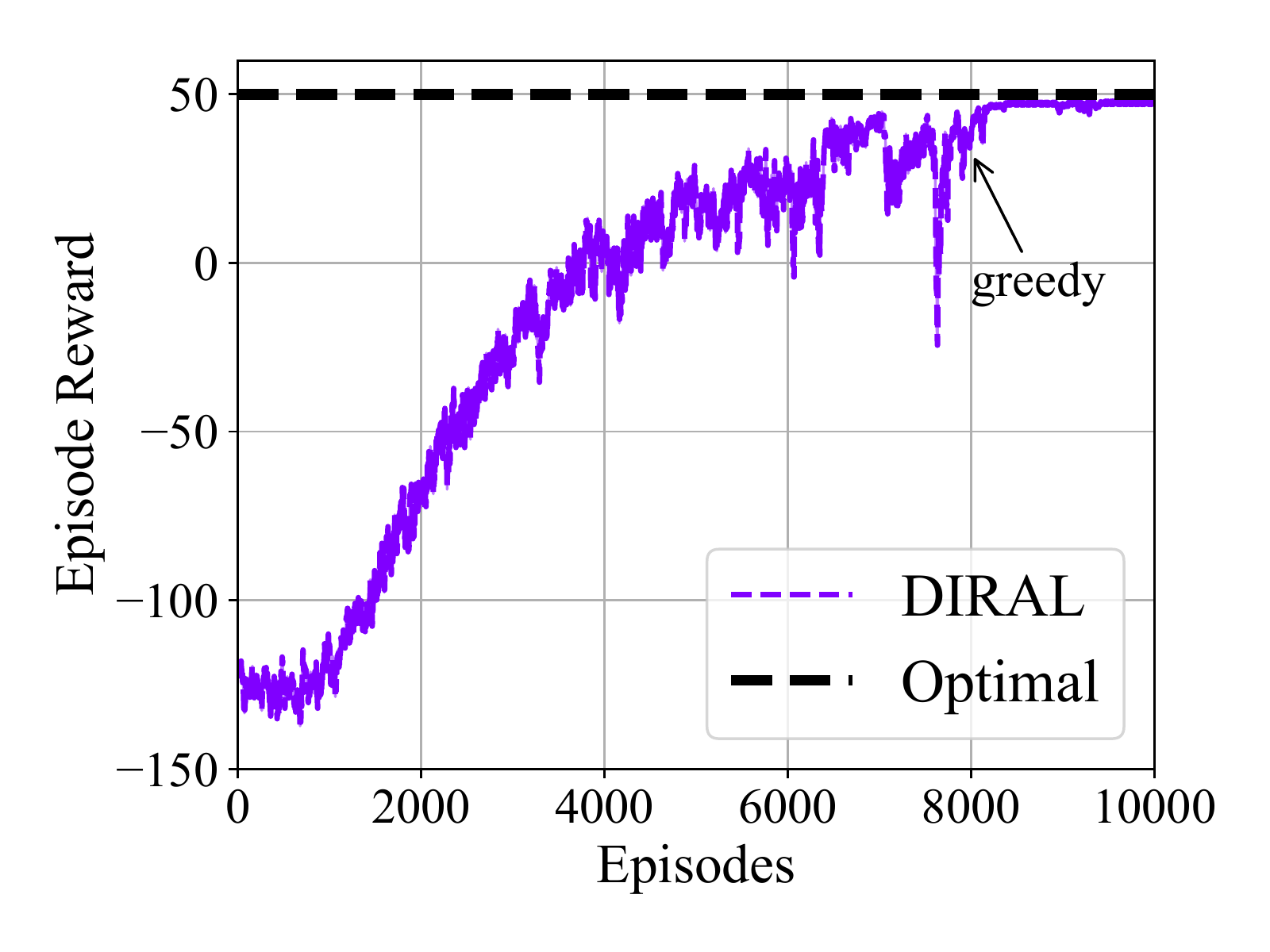}  
  \caption{Scenario 1: 4 vehicles, 3 resources}
  \label{fig:sub-firstt}
\end{subfigure}
\begin{subfigure}{.23\textwidth}
  \centering
  \includegraphics[width=\linewidth]{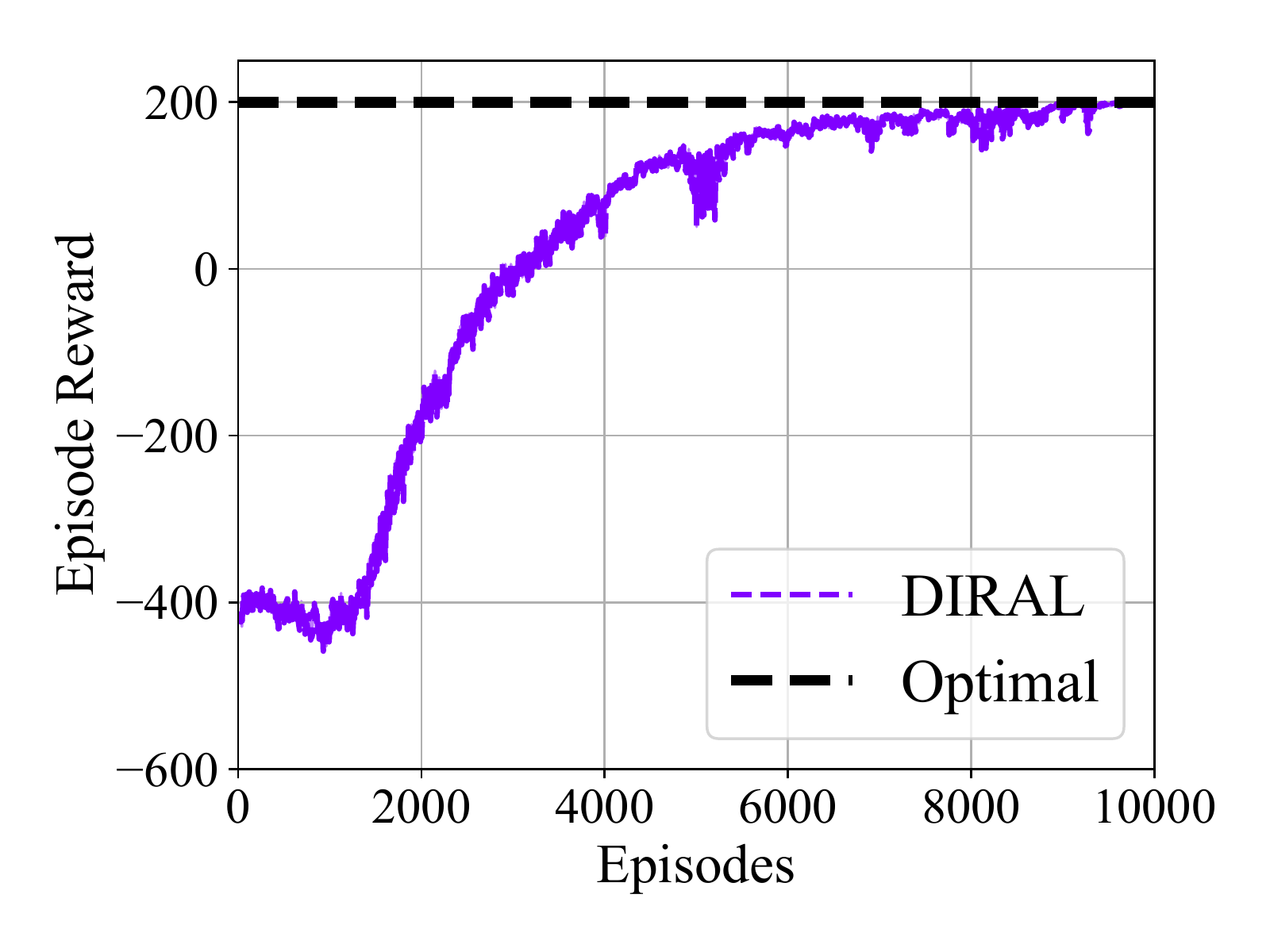}
  \caption{Scenario 4: 12 vehicles, 10 resources}
  \label{fig:sub-third}
\end{subfigure}
\vspace{-0.1cm}
\caption{Training performance of congested scenarios}
\vspace{-0.5cm}
\label{fig:train_per}
\end{figure}

We started the evaluations with a toy example as depicted in Figure~\ref{fig:exp_dist}, where we have $N=4$ vehicles and $K=3$ resources  at each time-step. The parameter to adjust the granularity of observations i.e. $B$ is set to $40$ in this toy example and $R$ value is selected proportional to the length of the highway e.g. $R=100$.
Each vehicle moves in the same direction with a unique velocity. We used the reward function in Equation \ref{eq:reward_design} with a slight modification for the scenarios $1$ and $2$. We gave neutral reward for the farthest vehicles that use the same resources to observe the desired behavior better. So, at each time-step the maximum reward of the system is 2 for the scenario $1$. The length of the each episode is 25 time-steps and we trained the model after each episode.  The convergence of this approach is illustrated in Figure \ref{fig:sub-firstt}. The proposed approach reaches the optimal policy for the considered scenario. 

We increased the granularity of observations to $B=100$ for large scenarios e.g. $3$, $4$, and $5$ to capture the mobility pattern of all the vehicles. Although higher $B$ increases the state space, DQN with LSTM architecture is able to find good estimates of Q-values. The model can differentiate the positions of vehicles better with increasing granularity of observations. We use the reward in Equation \ref{eq:reward_design} with $r_{reuse}=250m$ to motivate far vehicles to use the same resources. Note that, the proposed reward works also with large number of vehicles. As seen in \ref{fig:sub-third}, DIRAL is able to evolve more smoothly. This is mainly caused by the fact that we have a relatively simpler mobility model for the scenario $4$. 
We mainly aimed at proving convergence with the test simulator and due to space limitations we do not share all the training performances.  

\subsection{RealNeS Analysis}
The trained policies for the scenarios $2$, $3$, $4$, and $5$ in the test simulator are deployed to real time network simulator, and the performance of DIRAL is compared with random scheduling and SPS. SPS algorithm consists of \textit{sensing}, \textit{selection}, and \textit{reselection}. With \textit{sensing}, each vehicle monitors Received Signal Strength Indicator (RSSI) of shared resources for the last 1000 slots.  For \textit{selection}, resources with expected RSSI lower than a threshold form a resource pool and a resource is selected randomly from the shared pool. If the size of the resource pool is smaller than 20\% of all shared resources, then the threshold is increased by 3dB, and the \textit{selection} procedure is repeated. After the \textit{selection} process, a vehicle exploits the same resource for the subsequent $\sim[5, 15]$ transmissions that is set by the reselection counter. The reselection counter is decreased by one after every transmission and when it reaches zero, the vehicle continues to use the same resource with the probability of 0.8 or selects a new resource. In random scheduling, each vehicle randomly selects one resource among the shared resources in every transmission. 

\begin{figure}
 \centering
\begin{subfigure}{.49\textwidth}
    \includegraphics[width=1.1\linewidth]{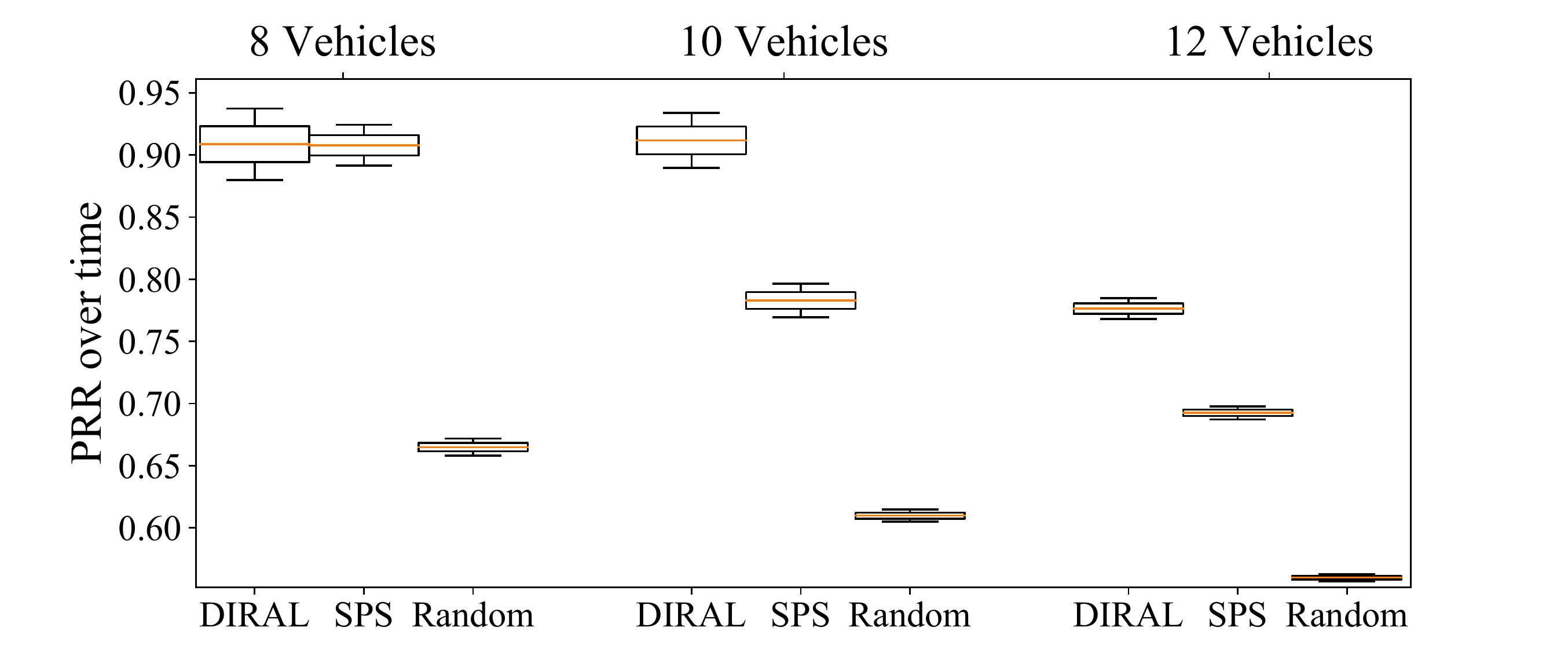}
    \caption{PRR of various vehicles with 10 resources.}
    \label{fig:10r_conf}
\end{subfigure}
\begin{subfigure}{.4\textwidth}
  \centering
  \includegraphics[width=\linewidth]{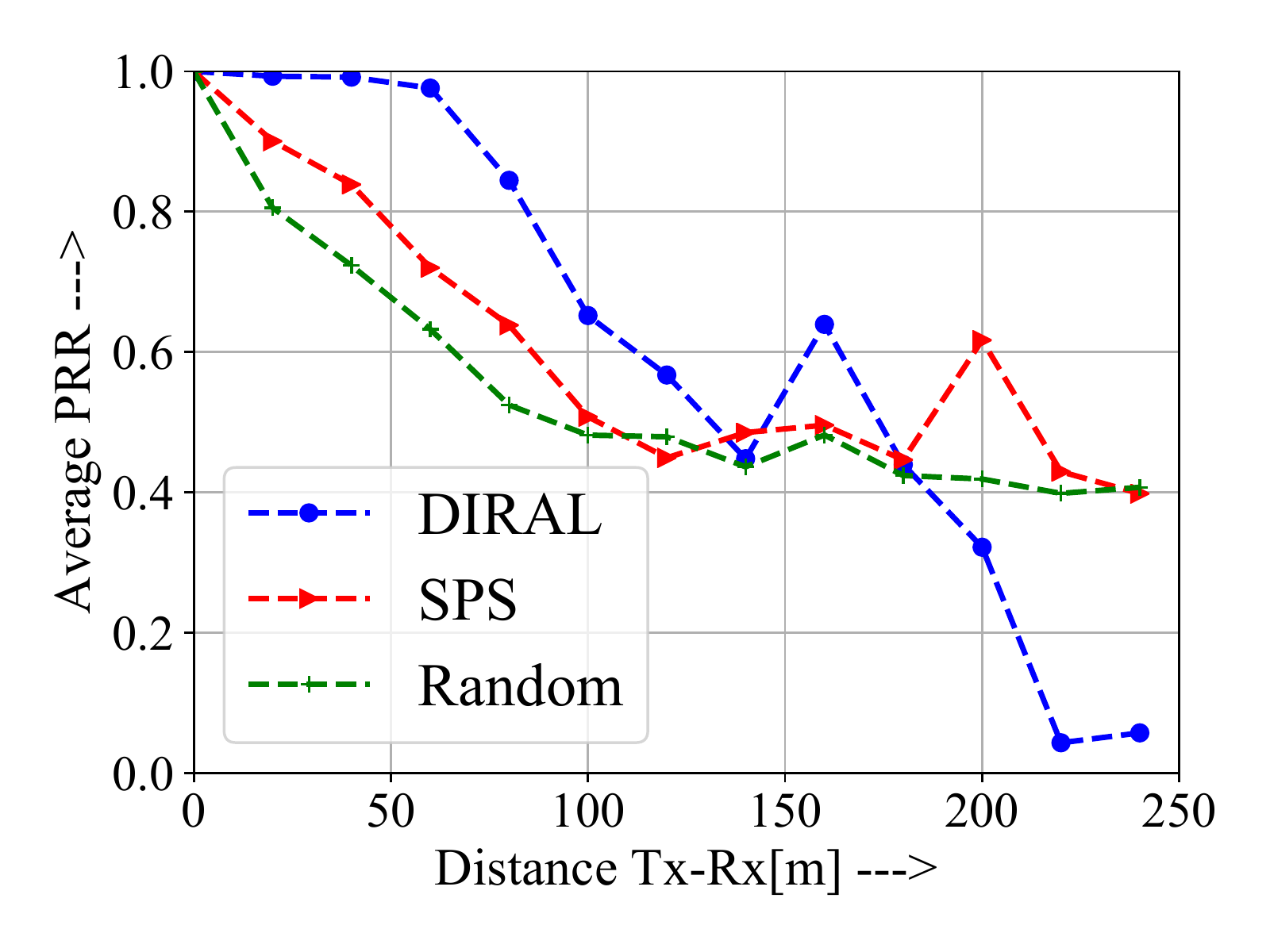}  
  \caption{Scenario 2}
  \label{fig:sub-third_prr}
\end{subfigure}

\caption{RealNeS analysis}
\label{fig:eval_per}
\end{figure}

We compare the performance of DIRAL for both congested and non-congested case. The measurements for evaluation are taken every 100 transmissions over 1000 seconds. The PRR values are illustrated with boxplots in Figure \ref{fig:10r_conf} for scenarios $3$, $4$ and $5$. This indicates that for a congested scenario DIRAL outperforms both SPS and random scheduling in terms of PRR. SPS algorithm is able to perform as well as DIRAL only when the number of resources higher than the number of vehicles. PRR as a function of the distance between transmitter and receiver is depicted in Figure~\ref{fig:sub-third_prr} for scenario 2. The system achieves very high PRR values for near vehicles compared to SPS and random while sacrificing communication with far vehicles by exploiting the same resources.


\section{Conclusion and Future Work} \label{sec:conc}
In this work, we proposed a novel algorithm, distributed resource allocation with multi-agent DRL (DIRAL) for out of coverage vehicular communications. The algorithm proposes a novel solution to the congestion problem that can rise naturally due to the uncontrolled mobility of the vehicles. We show that our proposal improves the PRR in a congested scenario by up to $20\% $. This algorithm enables V2X to be used more reliably for out of coverage scenarios. The practicality of our results are demonstrated by training in a simple Python based simulator and deploying on a more elaborate C++ based system simulator (RealNeS). 

The results presented herein are limited to vehicles moving in single direction. However, the case of vehicles moving in different directions can be handled by means of separate resource pools for different directions and applying DIRAL within the resource pools.  A solution with resource pools is left for future work. 

\bibliography{infocom21}
\bibliographystyle{IEEEtran}
\end{document}